\documentstyle[epsf,psfig]{mn}

\begin{document}
 \title[AX J0051$-$733]
     {Observations of the unusual counterpart to the X-ray pulsar AX J0051$-$733
  \thanks{Partially based on observations collected at the South African 
Astronomical Observatory and the European Southern Observatory, Chile 
(ESO 65.H-0314)}}

 \author[M.J.Coe et al.]
 {M.J. Coe$^{1}$, N.J. Haigh$^{1}$, S.G.T. Laycock$^{1}$,
I. Negueruela$^{2}$ \& C.R. Kaiser$^{1}$\\
$^{1}$ Physics and Astronomy Dept., The University, Southampton, SO17
 1BJ, UK. \\
$^{2}$Observatoire de Strasbourg, 11 rue de l'Universite, Strasbourg
 67000, France.\\
 }

 \date{Accepted \\
 Received : Version Nov 6 2001\\
 In original form ..}

 \maketitle

 \begin{abstract} 

We report optical and IR 
observations of the ASCA X-ray pulsar system AX J0051-733. The
relationship between the X-ray source and possible optical
counterparts is discussed. Long term optical data from over 7 years
are presented which reveal both a 1.4d modulation and an unusually
rapid change in this possible binary period. Various models are discussed.

 \end{abstract}


 \section{Introduction}

High Mass X-ray binaries (HMXBs) are traditionally divided into 
Be/X-ray and supergiant binary systems.  
 A survey of the literature reveals that of
 the 96 proposed massive X-ray binary pulsar systems, 67\% of the
 identified systems fall within the Be/X-ray group of binaries.  The
 orbit of the Be star and the compact object, a neutron star, is
 generally wide and eccentric.  The optical star exhibits H$\alpha$
 line emission and continuum free-free emission (revealed as excess
 flux in the IR) from a disk of circumstellar gas. Most of the Be/X-ray
 sources are also very transient in the emission of X-rays.

The source that is the subject of this paper, AX J0051-733, lies in
the Small Magellanic Cloud, a region of space that is extremely rich
in HMXBs. It was reported as a 323s pulsar by Yokogawa \& Koyama
(1998) and Imanishi et al (1999). Subsequently Cook (1998) identified
a 0.7d optically variable object within the ASCA X-ray error
circle. The system was discussed in the context of it being a normal
HMXB by Coe \& Orosz (2000) who presented some early OGLE data on the
object identified by Cook (1998) and modelled the system
parameters. Coe \& Orosz identified several problems with
understanding this system, primarily that if it was a binary then its
true period would be 1.4d and it would be an extremely compact
system. In addition, the combination of the pulse period and such a
binary period violates the Corbet relationship for such systems
(Corbet, 1986).

In this paper we report on extensive new data sets from both OGLE and
MACHO, as well as a detailed photometric study of the field. The
results reveal many complex observational features that are hard to
explain in the traditional Be/X-ray binary model.

\section {X-ray source location}

\begin{figure}
 \begin{center}
 \psfig{file=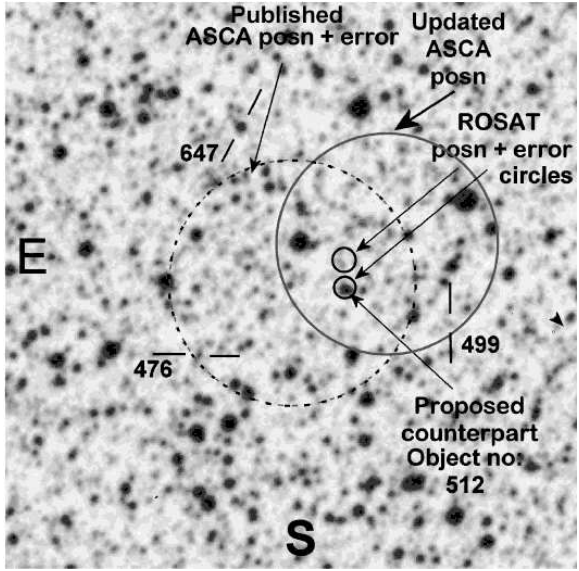,width=3in}
 \end{center}
 \caption{Finding charts for AX J0051-733 covering a field of 4 x 4
arcmin created using an 
optical V band image from this work. The northern ROSAT
circle refers to RX J0050.8-7316 and the southern one 
is that of RX J0050.7-7316. \label{optir}}
 \end{figure}

As will be seen from the photometric results presented below, it is critical
to establish the correct optical counterpart to the X-ray
pulsar. In particular, it is vital to clearly link the ASCA source to
an optical object, and other ROSAT X-ray sources may, or may not be
relevant (because no pulsations have been detected from ROSAT
objects). Figure~\ref{optir} illustrates the somewhat complex situation
associated with this object. In this figure the large dotted circle
indicates the original ASCA X-ray position and uncertainty from Imanishi
et al (1999). Within this error circle lie the much smaller error
circles of the ROSAT sources RX J0050.8-7316 (Cowley et al, 1997) and
RX J0050.7-7316 (Kahabka 1998). Subsequently, the position of the ASCA
error circle was refined to the large solid circle shown in the figure
(Imanishi 2001, private communication).

Within the ROSAT error circle for RX J0050.7-7316 and the revised ASCA
circle for AX J0051-733 lies an obvious optical object that has been
proposed as the counterpart to both of these X-ray objects (Cowley et
al, 1997, Schmidtke \& Cowley, 1998, Coe \& Orosz, 2000). It is a blue
star exhibiting variability which strongly suggests that it is a Be
star companion to the X-ray pulsar. This object also shows a
strong 0.7d optical modulation (or possibly twice that value) which
could be associated with a binary period of the system (Cook 1998, Coe
\& Orosz 2000). However, this period is very short for a HMXB and the
modulation signature atypical of that seen from such objects.

Consequently, it was felt necessary to revisit the linking of this
optical object with the ASCA pulsar to make sure that some other
candidate was not more appropiate within the X-ray error circle.

\section{Optical \& IR counterpart search}

Optical photometric observations were taken from the SAAO 1.0m
telescope on 2 October 1996. The data
were collected using the Tek8 CCD giving a field of $\sim$6 x 6
arcminutes and a pixel scale of 0.6 arcsec/pixel. Observations were
made through standard Johnson V \& R filters plus an H$\alpha$ filter. The
standard star E950 was used for photometric calibration. From these
CCD frames a R-H$\alpha$ colour index was created and this was plotted
against the V band flux for $\sim$800 objects.


On the assumption that our optical counterpart was likely to be a
H$\alpha$ bright system, all the objects in the top third of the
colour-magnitude plot were examined and their location in the field
identified. Only four such objects were determined to be in, or
close to, the ASCA error circle. These are numbered 476, 499, 512 and
647 in Figure~\ref{optir} 
(object no: 512 is the proposed counterpart
to the ROSAT sources). All the other objects with an R-H$\alpha$ index
$\ge-1.0$ lie well away from the region of interest.

The average $B$, $V$ \&$ I$ colours of these four objects were extracted from
the OGLE database and are presented in Table 1. In addition, IR
magnitudes for two of the objects are also presented that were
extracted from the 2MASS survey data base, the
other 2 candidates were too faint to be detected in that survey.

\begin{table}
\centering
\caption{Optical photometric values taken from the OGLE database and
IR values from the 2MASS survey.}
\begin{tabular}{cccccc}

ID & V & B-V & V-I & J & K \\
&&&&\\
476 & 18.70 & 1.00 & 1.08 & - & -\\
499 & 17.21 & -0.08 & 0.15 & - & -  \\
512 & 15.44 & -0.03 & 0.17 & 15.3 & 14.8 \\
647 & 15.69 & 0.07 & 0.26 & 16.5 & 15.9 \\
\end{tabular} 
\end{table} 

\begin{figure*}
 \begin{center}
 \psfig{file=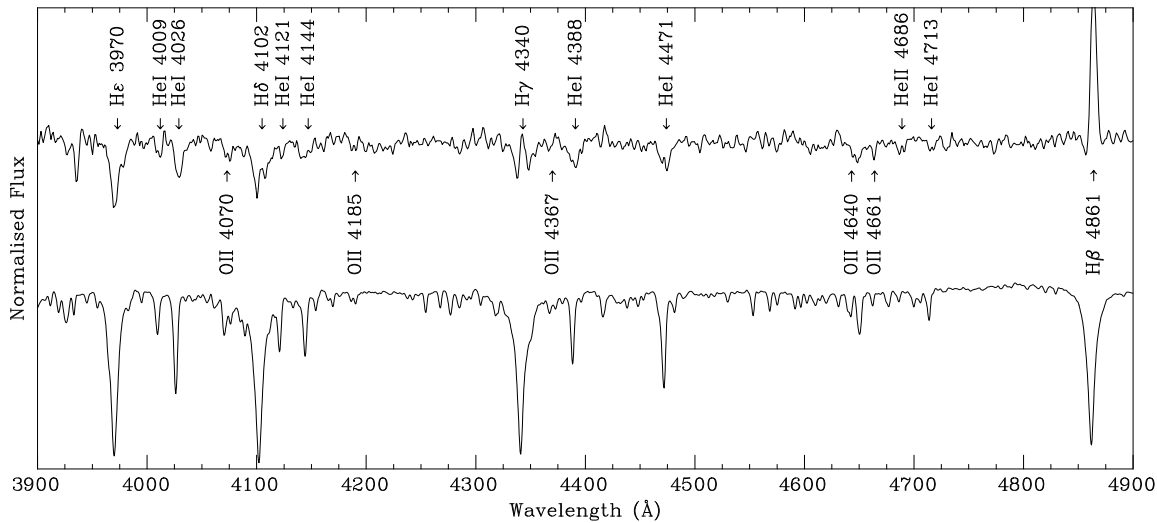,angle=-90,width=6in}
 \end{center}
 \caption{ Blue spectrum of Object 512 (upper spectrum) compared to a
B0.5V standard at a similar resolution. Note the presence of relatively
strong Na\,{\sc ii}~$\lambda 3934$\AA\ in the spectrum of Object~512, 
presumably of interstellar origin.
\label{os}}
 \end{figure*}

To confirm the nature of Object 512 as a B or Be star, optical spectra
were obtained on 3 occasions (1 Nov 1999, 15 Sep 2000 and 22 Oct 2000)
from the ESO 1.52-m telescope at La Silla Observatory, Chile, equipped
with the Boller \& Chivens spectrograph. The no: 33 holographic grating
was used, which gives a resolution of $\sim$1\AA/pixel. Since no
obvious variations were seen between the spectra they were combined to
increase the signal-to-noise ratio.  The resulting spectrum is
presented in Figure~\ref{os}. In this figure our spectrum is compared
to that of the B0.5V standard 40 Per. Object 512 is obviously a Be star, 
with H$\beta$ and H$\gamma$ in emission and most other lines affected by
emission components. The presence of weak He\,{\sc ii}~$\lambda$4686~\AA\ 
places the object close to B0V (Walborn \& Fitzpatrick 1990). Though several O\,{\sc ii} lines are 
present, C\,{\sc iv}~$\lambda$4650~\AA\ is surprisingly absent. The 
relatively weak Si\,{\sc iii} and Si\,{\sc iv} lines seen in 40 Per are
not easily detectable in object 512, which is compatible with the lower 
metallicity of the SMC, but unexpected in view of the rather strong 
O\,{\sc ii} lines.

\section{OGLE and MACHO data}

The field of AX J0051-733 lies within the areas covered by both the
OGLE 
\footnote [1] {http://sirius.astrouw.edu.pl/~ogle}
and MACHO 
\footnote [2] {http://wwwmacho.mcmaster.ca}
monitoring programmes. Hence excellent photometric coverage exists for
the brighter counterparts for a total of nearly 7 years. 

\begin{figure}
 \begin{center}
 \psfig{file=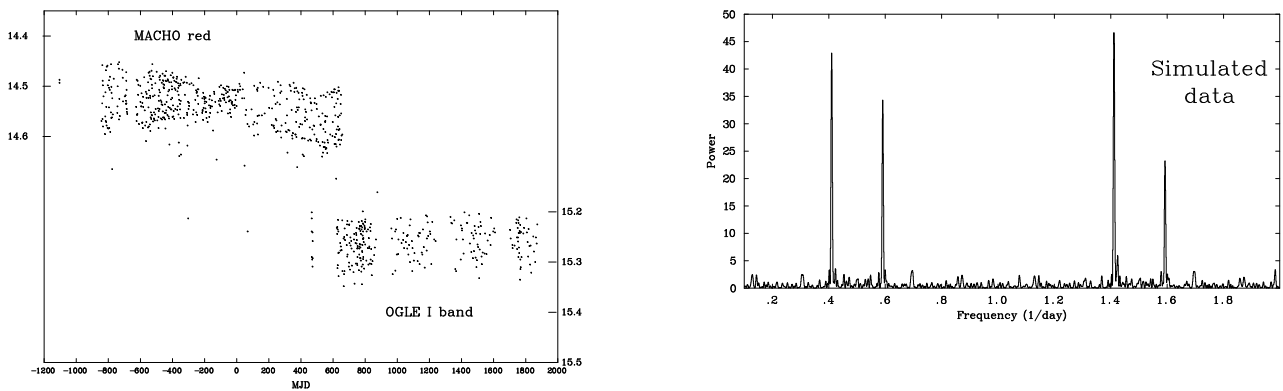,angle=-90,width=3in}
 \end{center}
 \caption{Approximately 7 years of photometric observations of the
proposed counterpart to AX J0051-733 taken from the MACHO and OGLE
data bases. The date axis has MJD = JD - 2450000. In both cases the
magnitude scale is indicated, though the MACHO one is described as
``approximately R''. \label{mo}}
 \end{figure}

Detailed $I$ band photometry was obtained from the OGLE data base for
objects numbered 499 (no significant variability), 647 (some evidence for
long term changes comparable to the length of the data set) and
512. As Cook (1998) and Coe \& Orosz (2000) have already shown from
subsets of the OGLE/MACHO data, this object exhibits a strong clear
sinusoidal modulation at $\sim0.7$d. The combined OGLE and MACHO data
set for this object is presented in Figure~\ref{mo}.

Though the precise modulation is not obvious from this figure, it
clearly shows the varying amplitude of the modulation over the data
set. If the total data set is analysed for periodic behaviour, then a
period of 0.70872d is determined using a Lomb-Scargle
analysis. However, this period is the average of the data, because if
one splits up the data set into $\sim150$d samples a slightly
different period is found for each one.

\begin{figure}
 \begin{center}
 \psfig{file=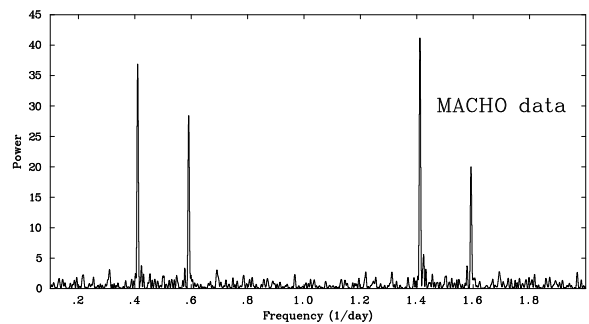,width=3in}
 \end{center}
 \caption{Comparison of a Lomb-Scargle power spectrum for a $\sim$150d
section of MACHO data (lower panel) and a simulated data set (upper
panel). The simulated data set consists of a pure sine wave with the
same window function as the raw data (see text for more details). \label{pow}}
 \end{figure}

Figure~\ref{pow} illustrates the Lomb-Scargle power spectrum for one such
subset of data. To check on the aliasing with the Nyquist frequency
and the effects of the window function, a simulated data set was
created. This data set consists of a single sinewave with period and
amplitude determined from the original data sampled with exactly the
same temporal structure as the original data. As can be seen by
comparison between the two power spectra in Figure 4, there is no
significant difference. Thus the conclusion is that there are no other
frequencies present in the original data set. 

\begin{figure}
 \begin{center}
 \psfig{file=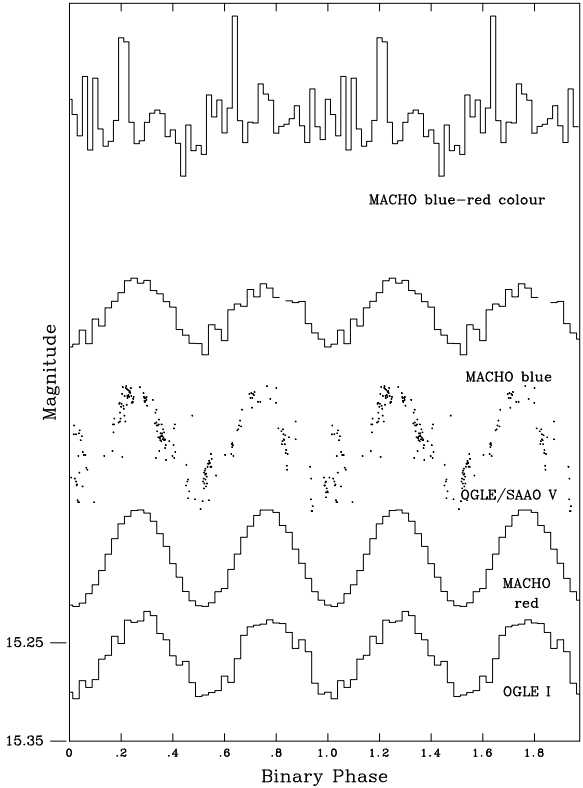,width=3in}
 \end{center}
 \caption{The lightcurves obtained by folding a $\sim$150d sample of
MACHO and OGLE data at the period of 1.4174d. Because the OGLE $V$
filter coverage is very sparse, data from several observations at SAAO in
this band have been added to the illustrated data set. The magnitude
scale on the left only refers to the OGLE $I$ band data, all the other
photometric bands have been arbitarily shifted upwards by a constant
amount to fit conveniently on the figure. In each case the data sets
have been phase shifted to coalign with the OGLE $I$ band data set (see
text for further details).
The uppermost curve shows the colour information
obtained from the same MACHO data set used to construct the light
curve in the figure. \label{lc}}

 \end{figure}

The shape of the modulation was determined by folding one of the MACHO
and OGLE data sets at the determined period for that data set. The
result of this is illustrated in Figure~\ref{lc}. Lightcurves from four
different filters are shown in this figure. In the case of the V band,
the OGLE data are rather sparse since this is not their main filter,
and so the data have been supplemented by observations taken over
several nights from the SAAO 1.0m in January 1999 and 2000. Since this
data set is not as uniform as the other 3 bands the individual
results are presented in the phase diagram rather than a folded light
curve. Overall, from this figure, the extremely sinusoidal nature of
the modulation is very clear.

The topmost curve in Figure~\ref{lc} is a measure of the colour of the object
obtained by subtracting individual MACHO blue measurments from their
red measurements taken on the same night. If the colour data are also
subjected to the same Lomb-Scargle analysis, then the same period
emerges from the data, but the depth of modulation is clearly very
small.

\begin{figure}
 \begin{center}
 \psfig{file=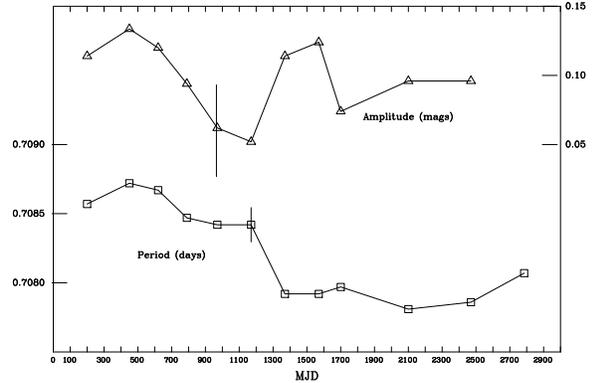,angle=-90,width=3in}
 \end{center}
 \caption{The lower curve shows the period history determined from the
combined MACHO and OGLE data sets. If the source is binary system then
we should expect the true binary period to be twice the value
indicated on the left hand axis. The upper curve shows the amplitude
of a sine wave fitted to each data block. In both cases a typical
error bar is indicated. The time axis has MJD = JD - 2449000. \label{per}}
 \end{figure}

Perhaps the most important result to emerge from the combined
OGLE/MACHO data set is the period history. Figure~\ref{per} illustrates this
by showing the periods determined from the 12 individual data
sets. From this figure one can clearly see that the period is changing
by a significant amount over the 7 years. The simplest interpretation
of the period change is a linear one, and from such an assumption a
value of 13.5s/year is found. More complex changes, for example a
sinusoidal modulation with a period of $\sim$3600d, may be speculated
upon but are not required by the data as yet.

The upper curve in Figure~\ref{per} shows the result of fitting a sine wave to
each individual data set and determining its amplitude. Though there
is clear evidence of amplitude variation by $\sim$40\% from both this
figure and Figure~\ref{mo}, it is not obvious that there is any significant
pattern to the change.

\section{Discussion}

\subsection{Optical candidate}

In trying to establish the optical counterpart to the ASCA pulsar one
must keep in mind that the ROSAT source is too weak to have shown any
detectable pulsations. Thus putting the ROSAT source aside for the
moment, the most objective approach is to just look at the
colour-magnitude diagram (Figure~\ref{cmd}). This diagram reveals just
two objects inside the best ASCA positional error circle - nos. 512
and 499. The other two objects lie in, or very close to the original
ASCA circle, but are now significantly less attractive as
counterparts. Object 499 is very faint compared to all other known
counterparts to HMXBs in the SMC, which typically have
$V\sim15-16$. Its colours and the presence of H$\alpha$ in
emission suggest a B3-4Ve star - a somewhat later spectral type than most
other Be/X-ray binary systems. 
Its OGLE lightcurve reveals nothing of interest and it is not a detectable IR
source in the 2MASS data. Hence it cannot be a strong contender for
the counterpart to AX J0051$-$733.

On the other hand, Object 512 has $V=15.4$ and a significant IR flux at
$J=15.3$. Both of these make it look like a classic counterpart to a
Be/X-ray binary system. If we compare this object to another SMC X-ray
pulsar system, 1WGA J0053.8$-$7226 (Buckley et al, 2001), we find it is
extremely similar. In 1WGA J0053.8$-$7226 we have $(B-V)=-0.06$ compared to
$-0.03$ in Object 512, and $(J-K)=0.62$ compared to 0.51 in Object
512. The $E(B-V)$ value found for many other SMC counterparts to
Be/X-ray systems is $\sim0.25$ (a combination of extinction to the SMC
plus local extinction due to cirumstellar material). Applying this to
the observed values for Object 512 given in Table 1 leads to an
identification for the spectral type of B0III-V.
Thus even before one considers the {\em ROSAT} source, one is led inexorably
to Object 512 being the prime candidate for the optical counterpart to
AX J0051-733. The presence of a convincing {\em ROSAT} source at the same
position adds significant extra weight to this conclusion.

The optical spectrum of Object 512 presented in Figure~\ref{os} is no
later and perhaps slightly earlier than the comparison standard. From
the colours presented in Table 1 and assuming $(B-V)_{0} = -0.26$ 
(Wegner 1994), this results in an extinction value
of $E(B-V)=0.23$, which confirms the number used above in interpreting
just the photometry. Assuming standard reddening, $A_{V} = 0.71$ and 
therefore, assuming a distance modulus to the SMC $(M-m)=18.9$, the absolute
magnitude for Object 512 is $M_{V} = -4.2$, which is in rather good
agreement with a spectral type in the B0-B0.5V range.

\subsection{Optical modulation}

The strong sinusoidal optical modulation of Object 512 is challenging
to interpret in terms of a traditional Be/X-ray binary model. Firstly, 
the expected
binary period of AX J0051$-$733 based on the Corbet diagram (Corbet,
1986) is 100--200d. Secondly, a binary period of just 1.4d involving a
Be star implies an extremely tight orbit -- the Keplerian orbital
radius would be $\sim14$ solar radii and the B0 star has a size of
$\sim8$ solar radii. Thirdly, if the period is really decreasing at
a rate of 13.5 s/year then this implies (Huang 1963) a mass transfer of
$10^{-5}\: M_{\odot}$/year for mass transfer between an 18$M_{\odot}$
Be star and a 1.4$M_{\odot}$ neutron star -- which is not only much
larger than that typically observed in HMXB systems ($\la 10^{-8}\: 
M_{\odot}$/year in most cases), but would also imply a much higher
X-ray luminosity unless the accretion mechanism is extremely inefficient
at converting gravitational potential into X-rays.

Mass transfer rates of this magnitude are deduced to exist in the EB
binary system $\beta$ Lyrae which is changing its $\sim13$d orbital
period at a rate of 19s/year (van Hamme, Wilson \& Guinan 1995). In
this case the change is to a longer period with the mass transferring
from the smaller B6-8 star to the more massive Be star. In our case,
the mass would be flowing in the opposite direction, i.e. from the
more massive object to a less massive one. The optical lightcurve of
$\beta$ Lyrae is similar to the one presented here for Object 512, but
with the notable difference that in $\beta$ Lyrae the two minima
are not of the same depth.

In fact the symmetry of the light curve is much more suggestive of
a W UMa type system. Unfortunately, the observed period of 1.4d
is much greater than any such reported system in the SMC (Rucinski
1997).  The maximum observed period is 0.8d and our period is well off
the end of the distribution. In addition, it is perhaps worth noting
that the predicted $(V-I)$ colour obtained from the distribution of such
objects and our binary period of 1.4d is $+0.026$, but from Table 1 it
can be seen that the observed $(V-I)$ for Object 512 is 0.17. Even
allowing for interstellar extinction this further adds to it being
unlikely that this system is of this class. 

The possibility of a blended variable star plus Be star can be
considered. For example, a chance superposition of Be star (to give the
observed colours) plus Cepheid or RR Lyrae (to give the optical
modulation). However, all of these models can be ruled out because of
either the magnitude of the period, or the depth of modulation, or the
shape of the lightcurve.

Interestingly the optical modulation is somewhat similar to the short
periodic modulation seen by Balona (1992) in Be stars in the cluster
NGC330 in the SMC. In this case Balona attributes this modulation to
surface features on the rapidly rotating objects. However, how the
period of such objects could change on a timescale of years is not
clear, unless the star is in a very wide binary system. It is possible
that the data in Figure~\ref{per} could be fitted to $\sim$10 year
sinusoidal modulation, but then the orbit of the neutron star would be
so distant from the Be star that it hard to see how accretion could
ever occur. In addition X-ray outbursts have been detected 3 times over
2 years from this system (Laycock, private communication) making such
a long orbit unlikely. 
Perhaps further optical data may clarify exactly what the shape of
the period change is on such timescales.

\subsection{A triple system?}

We are left with no convincing traditional scenario to explain all the
observational data. It is very hard to see how the orbital period
change seen in Figure~\ref{per} could possibly be caused by mass loss
from a normal B0 star at a rate of $10^{-5}$ $M_{\odot}$/year.  One
other possibility perhaps worth considering is that AX J0051$-$733 is a
triple system -- Be star plus another star in a tight 1.4d orbit, and
the neutron star in a highly eccentric 100$-$200d orbit around the
pair. Such a system could not only be intrinsically very stable since
most of the mass is concentrated in the inner binary pair, but the
transfer of angular momentum from the inner binary to the orbit of
the neutron star might also explain the evolution of the orbital
period.

Eggleton \& Kiseleva (1995) derive a critical parameter $X_{o}^{min}$
for a stable triple system, which is the period ratio between the outer
and inner orbits. For a system to be stable it is required that the
ratio of the orbital periods be greater than $X_{o}^{min}$. If we
assume that our inner 1.4d binary consists of a B0V star
(M=18$M_{\odot}$) and a 1$M_{\odot}$ star, while the third outer body
is a 1.4$M_{\odot}$ neutron star, then this parameter $X_{o}^{min}$ =
17. Assuming that the outer orbital period is actually given by the
position of AX J0051$-$733 on the Corbet diagram and has a value of
$\sim$100d, then this criterion is easily satisfied.

Bailyn \& Grindlay (1987) provide a formula for the rate of change of
size of the major axis of such a tight binary
(their Equation 7). Using their relationship, and assuming one of the
binary partners is the observed BO star, then it is possible to
predict the rate of change of orbital period as a function of the mass
of the other star in the inner binary. For masses of the order
15-20$M_{\odot}$ the predicted period change is $\sim$10 s/year.  This
number is in good agreement with the observed value of 13s/year and
suggests that the inner binary may, in fact, consist of two very
similar B-type stars. This, of course, would not present any problems
to the observed photometric or spectroscopic parameters of the
system. Even assuming that the two stars contribute equally to the
luminosity of the system would mean that their intrinsic magnitudes
are $M_{V}=-3.5$, still compatible with B0.5Ve. Furthermore, the
Eggleton \& Kiseleva criterion remains comfortably satisfied for such
a system.  The main problem raised by such scenario though would be
the very little space left between the two stars for a Be disk.

However, the evolution of this system would have to have been very
different from a classic Be/X-ray binary system. In particular, the
neutron star progenitor has probably evolved without any mass-transfer
to either of the objects in the inner binary. Consequently it must
have been much more massive in order to have reached its current state 
so long ahead of the other stars in the system. Clearly this solution for 
AX J0051$-$733
is also not without challenges.

\section{Conclusions}

Detailed optical observations and analysis have been carried out of
the proposed counterpart to AX J0051-733. The most likely counterpart
has been identified on the basis of its colours and H$\alpha$
emission. However this object is revealed to have a strong 0.7/1.4d
modulation from long-term MACHO and OGLE observations. Furthermore
this strong period is shown to be changing at a rate of 13.5s/year. It
is hard to reconcile all these observations with the classic Be/X-ray
binary model and further studies of this system are urgently required.

\section*{Acknowledgments}

We are extremely grateful to Andrzej Udalski and the OGLE team for
providing us with their data and to Kem Cook for help with the MACHO
data.  We are also grateful to the very helpful staff at the SAAO for
their support during these observations.  Many helpful comments from
Tom Marsh and Dave Buckley have been very much appreciated. All of the
data reduction was carried out on the Southampton Starlink node which
is funded by the PPARC.  NJH and SGTL are in receipt of a PPARC
studentships.

This paper utilizes public domain data obtained by the MACHO Project,
jointly funded by the US Department of Energy through the University
of California, Lawrence Livermore National Laboratory under contract
No. W-7405-Eng-48, by the National Science Foundation through the
Center for Particle Astrophysics of the University of California under
cooperative agreement AST-8809616, and by the Mount Stromlo and Siding
Spring Observatory, part of the Australian National University.

This publication makes use of data products from the Two Micron All
Sky Survey, which is a joint project of the University of
Massachusetts and the Infrared Processing and Analysis
Center/California Institute of Technology, funded by the National
Aeronautics and Space Administration and the National Science
Foundation.

We are also grateful to the referee, Lex Kaper, for several helpful
and interesting suggestions to improve the paper.

\bsp

\end{document}